\documentclass[11pt]{article}
\usepackage{latexsym,amsmath,amssymb,fullpage}

\newcommand{\nums}{{\mathbb N}}
\newcommand{\rats}{{\mathbb Q}}
\newcommand{\reals}{{\mathbb R}}
\newcommand{\nonneg}{{[0,\infty)}}
\newcommand{\two}{{\{0,1\}}}
\newcommand{\strs}{{\two^*}}
\newcommand{\prefeq}{\sqsubseteq}
\newcommand{\pref}{\sqsubset}
\newcommand{\success}[1]{{S^{\infty}[{#1}]}}
\newcommand{\RAND}{{\rm RAND}}
\newcommand{\seqs}{{\bf C}}
\newcommand{\tuple}[1]{{\langle{#1}\rangle}}
\newcommand{\map}[3]{{{#1}\;\colon\;{#2}\rightarrow {#3}}}

\newtheorem{theorem}{Theorem}[section]
\newtheorem{corollary}{Corollary}[section]
\newtheorem{definition}[theorem]{Definition}
\newenvironment{myproof}{\noindent{\bf Proof.}}{\hfill $\Box$}

\title{Gales and supergales are equivalent for defining constructive
Hausdorff dimension}

\author{Stephen A. Fenner\thanks{Computer Science and Engineering
Department, Columbia, SC 29208.  {\tt fenner@cse.sc.edu}}\\
University of South Carolina}

\date{August 16, 2002}

\begin{document}

\maketitle

\begin{abstract}
We show that for a wide range of probability measures, constructive
gales are interchangable with constructive supergales for defining
constructive Hausdorff dimension, thus generalizing a previous
independent result of Hitchcock \cite{Hitchcock:gales} and partially
answering an open question of Lutz \cite{Lutz:dimension2}.
\end{abstract}

\section{Introduction}

Various constructive, computable, and subcomputable versions of
classical Lebesgue measure have been proposed and studied
\cite{Lutz:measure} in order to quantify the sizes of complexity
classes (in the broadest sense of the word), and also to clarify the
idea of ``random sequence.''  In particular, the notion of
constructive measure and Martin-L\"of randomness has a rich history
\cite{MartinLoef:random,Schnorr:randomness}.

An useful idea related to Lebesgue measure is that of classical
Hausdorff dimension, which provides a fine-grained guage of the sizes
of null sets.  For example, Hausdorff dimension is a primary tool for
classifying geometric fractals \cite{Mandelbrot:fractals}.  As with
Lebesgue measure, various resource-bounded versions of Hausdorff
dimension have been studied \cite{Lutz:dimension1,Lutz:dimension2}.
In \cite{Lutz:dimension2}, Lutz concentrates on the constructive
Hausdorff dimension of individual sequences and (even) individual
finite strings.  Constructive dimension turns out to be closely
related to Kolmogorov complexity \cite{Mayordomo:dimension}, so
several results about the latter idea
\cite{Ryabko:dimension,Staiger,CaiHartmanis} inform the former.

Due to a result of Schnorr, constructive measure may be defined either
in terms of constructive martingales or in terms of constructive
supermartingales (see below for definitions).  Lutz defined
constructive dimension in terms of supergales for general computable
probability measures on the Cantor space \cite{Lutz:dimension2}, but
left open the question of whether this definition is equivalent to the
analogous one using gales (see below for definitions).  Hitchcock has
recently shown that for the uniform probability measure, gales and
supergales are indeed equivalent \cite{Hitchcock:gales}.  In the
current paper, we prove the equivalence for a wide range of
probability measures---those satisfying a certain reasonable balance
condition defined below.

Unfortunately, our techniques do not work for all computable
probability measures.  It is still an open question whether this
balance condition can be weakened or eliminated.

\section{Preliminaries}

We borrow most of our definitions and notation from
\cite{Lutz:dimension2}, which should be consulted for more details.
Let $\reals$, $\rats$, and $\nums$ be the set of real numbers,
rational numbers, and nonnegative integers, respectively.  Let $\strs$
be the set of finite binary strings, and let $\seqs$ be the set of
infinite binary sequences, i.e., the \emph{Cantor space}.  We let
$\lambda$ denote the empty string, and we let $|w|$ denote the length
of $w\in\strs$.  For any $x,y\in\strs\cup\seqs$ we write $x\prefeq y$
to mean $x$ is a prefix of $y$, and we write $x\pref y$ to mean $x$ is
a proper prefix of $y$.  A set $U\subseteq\strs$ is a \emph{prefix
set} if no element of $U$ is a proper prefix of any other element in
$U$.

A function $\map{d}{\strs}{\reals}$ is \emph{computable} if there is a
computable function $\map{d'}{\strs\times\nums}{\rats}$ such that for
all $x\in\strs$ and $r\in\nums$,
\[ \left|d'(x,r) - d(x)\right| \leq 2^{-r}. \]
A real number $s\in\reals$ is \emph{computable} if the constant
function $d(w) = s$ is computable.  A function
$\map{d}{\strs}{\reals}$ is \emph{lower semicomputable}, or
\emph{weakly computable} if there is a computable function
$\map{d'}{\strs\times\nums}{\rats}$ such that for all $x\in\strs$ and
$r\in\nums$,
\begin{enumerate}
\item $d'(x,r) \leq d'(x,r+1) < d(x)$ and
\item $d(x) = \lim_{r\rightarrow\infty} d'(x,r)$.
\end{enumerate}
Equivalently, $d$ is lower semicomputable iff the set $\{ (x,q) \in
\strs\times\rats \mid q < d(x) \}$ is computably enumerable (c.e.).
We can define upper semicomputability similarly, whence computability
is equivalent to upper and lower semicomputability combined.

A \emph{probability measure on $\seqs$} is a function
$\map{\nu}{\strs}{\nonneg}$ such that $\nu(\lambda) = 1$ and
\[ \nu(w) = \nu(w0) + \nu(w1) \]
for all $w\in\strs$.  The \emph{uniform probability measure} $\mu$ is
defined as $\mu(w) = 2^{-|w|}$.

Fix a probability measure $\nu$ and a real number $s\in\nonneg$.  A
\emph{$\nu$-$s$-supergale} is a function $\map{d}{\strs}{\nonneg}$
such that for all $w\in\strs$,
\begin{equation}\label{def:supergale}
\nu(w)^s d(w) \geq \nu(w0)^s d(w0) + \nu(w1)^s d(w1).
\end{equation}
A \emph{$\nu$-$s$-gale} is a $\nu$-$s$-supergale that satisfies
(\ref{def:supergale}) with equality.

An \emph{$s$-supergale} (respectively \emph{$s$-gale}) is a
$\mu$-$s$-supergale (respectively $\mu$-$s$-gale).

A \emph{$\nu$-supermartingale} (respectively \emph{$\nu$-martingale})
is a $\nu$-$1$-supergale (respectively $\nu$-$1$-gale).

A \emph{supermartingale} (respectively \emph{martingale}) is a
$\mu$-supermartingale (respectively $\mu$-martingale).

For example, an $s$-supergale $d$ satisfies
\[ d(w) \geq 2^{-s}[d(w0) + d(w1)]. \]

or any $\nu$-$s$-supergale $d$, we define its \emph{success set}
$\success{d} \subseteq\seqs$ by
\[ z \in \success{d}\; \mbox{ iff }\; \limsup_{w\pref z} d(w) =
\infty. \]
For $z\in\success{d}$ we say that \emph{$d$ succeeds on $z$}.  It is
well-known that a set $X\subseteq\seqs$ has Lebesgue measure zero iff
there is a martingale $d$ with $X\subseteq\success{d}$.

A $\nu$-$s$-supergale is \emph{constructive} if it is lower
semicomputable.

\begin{definition}\label{def:constructive-null}
Let $\nu$ be a probability measure on $\seqs$.  A set
$X\subseteq\seqs$ has \emph{constructive $\nu$-measure zero} if there
is a constructive $\nu$-martingale $d$ with $X\subseteq\success{d}$.
We say that $X$ has \emph{constructive $\nu$-measure one} if $\seqs -
X$ has constructive $\nu$-measure zero.  A sequence $R\in\seqs$ is
\emph{$\nu$-random} if $\{R\}$ does not have constructive
$\nu$-measure zero.  We let $\RAND_{\nu}$ be the set of all
$\nu$-random sequences.
\end{definition}

It is well-known that $\RAND_{\nu}$ has $\nu$-measure one for all
computable $\nu$ \cite{Lutz:dimension2}.  The most important case of
Definition~\ref{def:constructive-null} is when $\nu = \mu$.  The
following characterization of $\mu$-randomness was proved in
\cite{Schnorr:randomness}.  For a definition of Martin-L\"of
randomness, see \cite{MartinLoef:random} or
\cite{Schnorr:randomness}.

\begin{theorem}[Schnorr]
A sequence $R\in\seqs$ is $\mu$-random if and only if $R$ is random in
the sense of Martin-L\"of.
\end{theorem}

Schnorr also essentially proved that for computable $\nu$,
Definition~\ref{def:constructive-null} does not change if we replace
``$\nu$-martingale'' with ``$\nu$-supermartingale.''  That is,

\begin{theorem}[Schnorr
\cite{Schnorr:randomness,Schnorr:probability,Lutz:dimension2}]
\label{thm:equivalence}
Let $\nu$ be a computable probability measure on $\seqs$.  A set
$X\subseteq\seqs$ has \emph{constructive $\nu$-measure zero} if and
only if there is a constructive $\nu$-supermartingale $d$ with
$X\subseteq\success{d}$.
\end{theorem}

Lutz \cite{Lutz:dimension1,Lutz:dimension2} develops constructive
Hausdorff dimension as an analog both to classical Hausdorff dimension
and to constructive measure.

\begin{definition}\label{def:dimension}
Let $\nu$ be a probability measure on $\seqs$.  Let $X\subseteq\seqs$
be a set of sequences, and let
\[ {\cal G}(X) = \{ s\in\nonneg \mid \mbox{there is a constructive
$\nu$-$s$-supergale $d$ with $X\subseteq\success{d}$} \}. \]
Then $\inf {\cal G}(X)$ is the \emph{constructive $\nu$-dimension of
$X$} and is written $\dim_{\nu}(X)$.
\end{definition}

It is easy to show that $\dim_{\nu}(X)$ is always at most one, and if
$\dim_{\nu}(X) < 1$, then $X$ has constructive $\nu$-measure zero.

\section{Main Result}

Can we alternatively define constructive $\nu$-dimension as in
Definition~\ref{def:dimension} replacing ``$\nu$-$s$-supergale'' with
$\nu$-$s$-gale?  The proof of Theorem~\ref{thm:equivalence} does not
generalize to $s<1$.  We show, however, that for certain $\nu$
(including $\mu$), one can in fact make the replacement above in
Definition~\ref{def:dimension} without changing it.  To do this we
prove a weaker analog of Theorem~\ref{thm:equivalence} for $s<1$
(Theorem~\ref{thm:main}, below).  The case for $\nu = \mu$ was shown
by Hitchcock \cite{Hitchcock:gales}.  Our more general proof has some
elements similar to his, even though it was arrived at independently.
Our result does not hold for all computable $\nu$; we need the
following definition.

\begin{definition}
Let $\nu$ be a computable probability measure on $\seqs$.  We say that
$\nu$ is \emph{well-balanced} if there are constants $0 < \alpha < 1$
and $C>0$ such that for all $w\in\strs$,
\[ 0 < \nu(w) \leq C\alpha^{|w|}. \]
\end{definition}

Note that the uniform measure $\mu$ is well-balanced ($\alpha =
\frac{1}{2}$ and $C=1$).  More generally, if
\[ \liminf_{w\in\strs,\;b\in\two} \frac{\nu(wb)}{\nu(w)} > 0, \]
then $\nu$ is well-balanced (but not conversely).

The following theorem, which should be compared with
Theorem~\ref{thm:equivalence}, immediately implies that
$\nu$-$s$-gales are equivalent to $\nu$-$s$-supergales for defining
constructive Hausdorff $\nu$-dimension, for all well-balanced $\nu$.

\begin{theorem}\label{thm:main}
Let $\nu$ be a well-balanced computable probability measure on
$\seqs$, and let $s\in\nonneg$.  For every constructive
$\nu$-$s$-supergale $d$ and every computable $s'>s$, there is a
$\nu$-$s'$-gale $d'$ such that $\success{d} \subseteq \success{d'}$.
\end{theorem}

\begin{myproof}
We generalize an argument made in \cite{Schnorr:randomness} about
martingales.  Let $\nu$ be as in the statement of the theorem.  For
arbitrary $U \subseteq\strs$ and $t>0$ define
\begin{equation}\label{eqn:maindef}
d_U^t(w) = \frac{1}{\nu(w)^t}\left(\sum_{u\;\colon\; wu\in U}\nu(wu)^t +
\sum_{n<|w| \;\colon\; w[0..(n-1)]\in U}
\frac{\nu(w[0..(n-1)])^t}{2^{|w|-n}}\right).
\end{equation}
This definition makes sense provided the first sum on the right-hand
side converges.  Clearly, this will be true for all $w$ if it is true
for $w = \lambda$, i.e, if
\[ d_U^t(\lambda) = \sum_{u\in U}\nu(u)^t < \infty. \]
Assume that $d_U^t(\lambda)$ is indeed bounded.  It then follows that
$d_U^t$ is a $\nu$-$t$-gale.  This can be seen as follows: if $U$ is a
prefix set, then we may argue as in \cite{Schnorr:randomness}---at
most one term on the right-hand side of (\ref{eqn:maindef}) is
nonzero, and so we have two cases: some prefix of $w$ is in $U$, or
otherwise.  The equation for a $\nu$-$t$-gale is easy to check in
either case.  Now an arbitrary $U$ (not necessarily a prefix set) can
be partitioned into the union $U = V_0 \cup V_1 \cup V_2 \cup \cdots$
of pairwise disjoint prefix sets
\[ V_i = \{w\in U \mid \mbox{exactly $i$ many proper prefixes of $w$
are in $U$} \}, \]
and it is then clear from (\ref{eqn:maindef}) that
\[ d_U^t = d_{V_0}^t + d_{V_1}^t + d_{V_2}^t + \cdots. \]
Thus $d_U^t$ is the sum of $\nu$-$t$-gales and so is a $\nu$-$t$-gale.
Note that if $t$ is computable and $U$ is c.e., then $d_U^t$ is
constructive.

Let $0\leq s<s'$ and $d$ be as in the statement of the theorem, and
let $c,\epsilon > 0$ be such that $\nu(x) \leq 2^{c-\epsilon|x|}$ for
every $x\in\strs$.  Such $c$ and $\epsilon$ exist because $\nu$ is
well-balanced.  We may assume without loss of generality that
$d(\lambda) \leq 1$.  For any $k\in\nums$ we then have
\[ \sum_{w \in \two^k} d(w)\nu(w)^s \leq d(\lambda) \leq 1 \]
by Lemma 3.3 in \cite{Lutz:dimension2}.  For each $i\in\nums$ let
\[ U_i = \{ w\in\strs \mid d(w) > 2^i \}. \]
Then for all $i,k\in\nums$,
\[ \sum_{w\in U_i\cap \two^k} \nu(w)^s \leq 2^{-i}
\sum_{w\in\two^k} d(w)\nu(w)^s \leq 2^{-i}, \]
and hence,
\begin{eqnarray*}
\sum_{w\in U_i\cap \two^k} \nu(w)^{s'} & = &
\sum_{w\in U_i\cap \two^k} \nu(w)^s \nu(w)^{s'-s} \\
& \leq & 2^{(s'-s)(c-\epsilon k)} \sum_{w\in U_i\cap \two^k} \nu(w)^s \\
& \leq & 2^{(s'-s)(c-\epsilon k) - i}.
\end{eqnarray*}
This in turn yields, for all $i\in\nums$,
\begin{eqnarray*}
d_{U_i}^{s'}(\lambda) & = & \sum_{w\in U_i} \nu(w)^{s'} \\
& = & \sum_{k\in\nums} \sum_{w\in U_i\cap \two^k} \nu(w)^{s'} \\
& \leq & \sum_{k\in\nums} 2^{(s'-s)(c-\epsilon k) - i} \\
& = & 2^{-i} \frac{2^{(s'-s)c}}{1-2^{-(s'-s)\epsilon}} \\
& < & \infty,
\end{eqnarray*}
which means that $d_{U_i}^{s'}$ is well-defined.

Finally we define, as in \cite{Schnorr:randomness},
\begin{equation}\label{eqn:dprime}
d' = \sum_{i\in\nums} i\cdot d_{U_i}^{s'}.
\end{equation}
We have $d'(\lambda) \leq C\cdot\sum_{i\in\nums} i2^{-i} < \infty$
($C$ is a positive constant).  Being the sum of $\nu$-$s'$-gales, $d'$
itself is a $\nu$-$s'$-gale, and since the set $U = \{ \tuple{i,w}
\mid w\in U_i \}$ is clearly c.e., $d'$ is constructive.  For any
$z\in\seqs$, suppose $z\in \success{d}$.  Then for each $i\in\nums$
there is a prefix $w_i$ of $z$ such that $d(w_i) \geq 2^i$, i.e.,
$w_i\in U_i$.  But then by (\ref{eqn:maindef}) we have
$d_{U_i}^{s'}(w_i) \geq 1$, whence $d'(w_i) \geq i$.  Thus $d'$
succeeds on $z$.
\end{myproof}

\bigskip

\noindent {\bf Remark.}  In the special case where $\nu = \mu$,
(\ref{eqn:dprime}) reduces to
\[ d'(w) = \sum_{i\in\nums} i \left( \sum_{u\;\colon\;wu\in U_i}
2^{-s'|u|} + \sum_{n<|w|\;\colon\;w[0..(n-1)]\in U_i}
2^{(s'-1)(|w|-n)} \right). \]

\begin{corollary}
Let $\nu$ be a well-balanced computable probability measure on
$\seqs$.  Let $X\subseteq\seqs$ be a set of sequences, and let
\[ {\cal G}(X) = \{ s\in\nonneg \mid \mbox{there is a constructive
$\nu$-$s$-gale $d$ with $X\subseteq\success{d}$} \}. \]
Then $\inf {\cal G}(X) = \dim_{\nu}(X)$.
\end{corollary}

\begin{corollary}[Hitchcock \cite{Hitchcock:gales}]
Let $X\subseteq\seqs$ be a set of sequences, and let
\[ {\cal G}(X) = \{ s\in\nonneg \mid \mbox{there is a constructive
$s$-gale $d$ with $X\subseteq\success{d}$} \}. \]
Then $\inf {\cal G}(X) = \dim_{\mu}(X)$.
\end{corollary}

\section{Further Research}

If $\nu$ is not well-balanced, then the $d'$ defined in the proof of
Theorem~\ref{thm:main} may not exist.  There are clearly some
(ill-balanced) $\nu$ and constructive $\nu$-$s$-supergales $d$ such
that the first sum in (\ref{eqn:maindef}) with $U=U_i$ and $t=s'$ is
unbounded.  Perhaps one can restrict the sets $U_i$ in some way to
bound the sum.

If not, perhaps there is a condition on $\nu$ that is strictly weaker
than being well-balanced but still suffices to prove the theorem.  One
possible candidate is the following: there is an $\epsilon > 0$ such
that $\success{f} = \emptyset$, where $f$ is the $\epsilon$-gale
defined by $f(w) = 2^{\epsilon|w|}\nu(w)$.  If this is the case (and
if $\nu$ is computable and $>0$), then we'll say that $\nu$ is
\emph{weakly balanced}.  Clearly, well-balanced implies weakly
balanced, but the converse does not hold.  Does Theorem~\ref{thm:main}
hold for all weakly balanced $\nu$?


\end{document}